\documentstyle[12pt,aps,prd,preprint]{revtex}
\begin{document}
\title{Universal Upper Bound to the Entropy of a Charged System}
\author{Shahar Hod}
\address{The Racah Institute for Physics, The
Hebrew University, Jerusalem 91904, Israel}
\date{\today}
\maketitle

\begin{abstract}

We derive a {\it universal upper bound} to the entropy of a {\it charged}
system. The entropy bound follows from application of the generalized
second law of thermodynamics to a gedanken experiment in which an 
entropy-bearing charged system falls into a charged black
hole. This bound is {\it stronger} than the Bekenstein entropy
bound for neutral systems.
\end{abstract}
\bigskip

Black-hole physics mirrors thermodynamics in many 
respects \cite{Chris,ChrisRuf,Haw1,Haw2,Car,BarCarHaw}. 
According to the thermodynamical analogy in black-hole physics, the
entropy of a black hole \cite{Beken1,Beken2,Beken3} is given 
by $S_{bh}=A/4 \hbar$, where $A$ is the
black-hole surface area. (We use gravitational units in which $G=c=1$).
Moreover, it is widely believed that a system
consisting of ordinary matter interacting with a black hole will obey
the {\it generalized second law of thermodynamics} (GSL): 
``{\it The sum of the black-hole entropy and the common (ordinary) 
entropy in the black-hole exterior never decreases}''. This assumption
plays a fundamental role in black-hole physics.

In a {\it classical} context, a basic physical mechanism is known 
by which a violation of the GSL can be achieved:
Consider a box filled with matter of proper energy $E$ and 
entropy $S$ which is dropped into a black hole. The energy delivered
to the black hole can be arbitrarily {\it red-shifted} by letting the
assimilation point approach the black-hole horizon. As shown by
Bekenstein \cite{Beken3,Beken4}, if the box is deposited with no radial momentum
 a proper distance $R$ above the horizon, and then allowed to fall in such that

\begin{equation}\label{Eq1}
R < \hbar S/2 \pi E\  ,
\end{equation}
then the black-hole area increase (or equivalently, the increase in 
black-hole entropy) is not large enough to compensate for the decrease
of $S$ in common (ordinary) entropy. Arguing from the GSL, Bekenstein
\cite{Beken5} has proposed the existence of a universal upper bound on the
entropy $S$ of any system of total energy $E$ and effective proper radius $R$:

\begin{equation}\label{Eq2}
S \leq 2\pi RE/\hbar\  ,
\end{equation}
where $R$ is defined in terms of the
area $A$ of the spherical surface which circumscribe the system
\cite{Beken5} $R=(A/4\pi)^{1/2}$.
This restriction is {\it necessary} for enforcement of the GSL; the
box's entropy disappears but an increase in black-hole entropy occurs
which ensures that the GSL is respected provided $S$ is bounded 
as in Eq. (\ref{Eq2}). Evidently, this universal upper bound is a {\it
  quantum} phenomena (the upper bound goes to infinity as $\hbar \to
0$). This provides a striking illustration of the fact that the GSL is
intrinsically a {\it quantum} law. The universal upper bound
Eq. (\ref{Eq2}) has the status of a supplement to the second law; the
latter only states that the entropy of a closed system tends to a
maximum without saying how large that should be.

Other derivations of the universal upper 
bound Eq. (\ref{Eq2}) which are based on black-hole physics have been
given in \cite{Zas1,Zas2,Zas3,LiLiu}. Few pieces of evidence exist concerning the
validity of the bound for self-gravitating systems
\cite{Zas1,Zas2,Sorkin,Zurek}. However, the universal bound Eq. (\ref{Eq2})
is known to be true independently of black-hole physics for a variety
of systems in which gravity is 
negligible \cite{Beken6,Beken7,BekenSchi,SchiBeken,BekenGuen}.

In this paper we challenge the validity of the GSL 
in a gedanken experiment in which an entropy-bearing {\it charged} 
system falls into a charged black hole. We show that while
the upper bound Eq. (\ref{Eq2}) is a necessary condition for the 
fulfillment of the GSL, it is {\it not} a sufficient one.

It is not difficult to see why a {\it stronger} upper bound on 
the entropy of an arbitrary charged system
must exist: The electromagnetic interaction experienced 
by a charged body (which, of-coarse, was not
relevant in Bekenstein's gedanken experiment) can {\it decrease} 
the change in black-hole entropy (area). Hence, the GSL would be
violated unless the entropy of the charged system (what disappears from the
black-hole exterior) is restricted by a bound {\it stronger} 
than Eq. (\ref{Eq2}).

Furthermore, there is one disturbing feature of the universal 
bound Eq. (\ref{Eq2}). As was pointed out by Bekenstein \cite{Beken5} 
black holes conform to the bound; however, the
Schwarzschild black hole is the only black hole which 
actually {\it attains} the bound. This uniqueness of
the Schwarzschild black hole (in the sense that it is the {\it only} 
black hole which have the maximum entropy allowed by quantum theory and
general relativity) is somewhat 
disturbing. Recently, Hod \cite{Hod1} derived an
(improved) upper bound to the entropy of a {\it spinning} system and proved 
that {\it all} electrically neutral Kerr black holes 
have the {\it maximum} entropy allowed by quantum theory 
and general relativity.
Clearly, the unity of physics demands a stronger bound for
{\it charged} systems in general, and for black holes in particular.

In fact, the plausible existence of an upper bound stronger than
Eq. (\ref{Eq2}) on the entropy of a charged system 
has nothing to do with black-hole physics; a part of the energy of the
electromagnetic field residing outside the charged system seems to be irrelevant 
for the system's statistical properties. This reduce the phase space
available to the components of a charged system. Evidently, an improved
upper bound to the entropy of a charged system must {\it decrease}
with the (absolute) value of the system's charge. 
However, our simple argument cannot yield the exact dependence of the
entropy bound on the system's parameters: its energy, charge, and proper radius. 

In fact, black-hole physics (more precisely, the GSL) yields a
concrete expression for the universal upper bound.
Arguing from the GSL, we derive a universal upper bound to 
the entropy of a charged system which is {\it stronger} than the bound
Eq. (\ref{Eq2}). We consider a charged body (assumed to be spherical
for simplicity) of rest mass $\mu$, charge
$q$, and proper radius $b$, which is dropped into a (charged)
Reissner-Nordstr\"om black hole.

The external gravitational field of a spherically symmetric object of 
mass $M$ and charge $Q$ is given by the Reissner-Nordstr\"om metric

\begin{equation}\label{Eq3}
ds^2=-\left( {1-{{2M} \over r}+{{Q^2} \over {r^2}}} \right)dt^2+\left( {1-{{2M}
\over r}+{{Q^2} \over {r^2}}} \right)^{-1}dr^2+r^2d\Omega ^2\  .
\end{equation}
The black-hole (event and inner) horizons are located at

\begin{equation}\label{Eq4}
r_{ \pm}=M \pm (M^2-Q^2)^{1/2}\  .
\end{equation}

The equation of motion of a charged body on the Reissner-Nordstr\"om
background is a quadratic equation
for the conserved energy $E$ (energy-at-infinity) of the body \cite{Carter}

\begin{equation}\label{Eq5}
r^4 E^2 -2qQr^3E+ q^2Q^2r^2-
\Delta(\mu^2 r^2 +{p_{\phi}}^2)- (\Delta p_r)^2=0\  ,
\end{equation}
where $\Delta=r^2-2Mr+Q^2=(r-r_{-})(r-r_{+})$.
The quantities $p_{\phi}$ and $p_r$ are the conserved 
angular momentum of the body and its
covariant radial momentum, respectively. 

The conserved energy $E$ of a body having a radial turning 
point at $r=r_{+}+ \xi$ \cite{note1} (where $\xi \ll
r_{+}$) is given by Eq. (\ref{Eq5})

\begin{eqnarray}\label{Eq6}
E&=&{qQ \over r_{+}}+ {{\sqrt{\mu^2r_{+}^2 +p_{\phi}^2} 
(r_{+}-r_{-})^{1/2}} \over
  {r_{+}^2}} \xi^{1/2}\left\{1+
O\left[{\xi /(r_{+}-r_{-})}\right] \right\} \nonumber \\
&& -{qQ \over r_{+}^2} \xi \left [1+O({\xi /r_{+}}) \right]\  .
\end{eqnarray}
This expression is actually the effective potential (gravitational
plus electromagnetic plus centrifugal) for given 
values of $\mu, q$ and $p_{\phi}$.
It is clear that it can be {\it minimized} by taking $p_{\phi}=0$ (which
also minimize the increase in the black-hole surface area. This is
also the case for neutral bodies \cite{Beken3}).

In order to find the change in black-hole surface area caused by
an assimilation of the body, one should evaluate $E$ [given
by Eq. (\ref{Eq6})] at the point of capture, a proper distance $b$
outside the horizon. Thus, we should
evaluate $E$ at $r=r_{+}+ \delta (b)$, where $\delta(b)$ is 
determined by 

\begin{equation}\label{Eq7}
\int_{r_{+}}^{r_{+}+ \delta (b)} (g_{rr})^{1/2} dr = b\  ,
\end{equation}
where $g_{rr}=r^2/\Delta$. Integrating Eq. (\ref{Eq7}) one finds (for $b \ll r_{+}$)

\begin{equation}\label{Eq8}
\delta (b)=(r_{+}-r_{-}) {b^2 \over {4{r_{+}}^2}}\  .
\end{equation}

An assimilation of the charged body results in a change $dM=E$ in the
black-hole mass and a change $dQ=q$ in its charge.
Taking cognizance of Eq. (\ref{Eq6}) and using the first-law of 
black-hole thermodynamics

\begin{equation}\label{Eq9}
dM={\kappa \over {8\pi}} dA + \Phi dQ\  ,
\end{equation}
where $\kappa=(r_{+}-r_{-})/2r_+^2$
and $\Phi=Q/r_{+}$ are the
surface gravity ($2\pi$ times the Hawking temperature \cite{Haw3}) 
and electric potential of the black hole, respectively, one finds

\begin{equation}\label{Eq10}
(\Delta \alpha)_{min}={4\mu r_{+}
\over {(r_{+}-r_{-})^{1/2}}}
\delta(b)^{1/2} -
{4qQ \over {r_{+}-r_{-}}} \delta(b)\  ,
\end{equation}
where the ``rationalized area'' $\alpha$ is related to the black-hole
surface area $A$ by $\alpha = A/4 \pi$.
With Eq. (\ref{Eq8}) for $\delta(b)$ we find

\begin{equation}\label{Eq11}
(\Delta \alpha)_{min}(\mu,q,b,s)=2\mu b-{{qQb^2} \over {r_{+}^2}}\  ,
\end{equation}
which is the {\it minimal} black-hole area increase \cite{note2} for given values of the
body's parameters $\mu, q$ and $b$ [and for given black-hole parameters
$r_+$ and $Q$ ($s$ stands for these two parameters)]. 

Obviously the increase in black-hole surface area Eq. (\ref{Eq11})
can be {\it minimized} (for given values of the body's parameters) by
maximizing the black-hole electric field (given by $Q/{r_+}^2$). 
However, we must consider an
external electric field with a limited strength in order to keep it from
deforming and breaking the charged body. Evidently, a charged body
does not break up under its {\it own} electric field. Clearly, this
value of the field is very {\it conservative}; most bodies can be
subjected to much stronger electric fields without being
broken. However, our goal is to derive a
{\it universal} upper bound which is valid for {\it each and every}
charged system in nature, regardless of its specific internal 
structure (and regardless of its internal constituents). Hence, we
must consider an electric-field strength of this order of magnitude 
(this assures us that the charged body does not break up under the
external electric field).
Therefore, we find

\begin{equation}\label{Eq12}
(\Delta \alpha)_{min}(\mu,q,b)=2\mu b-q^2\  ,
\end{equation}
which is the minimal area increase for given values of the
body's parameters $\mu, q$ and $b$.

It is in order to emphasize the assumptions made in obtaining 
Eq. (\ref{Eq12}). By keeping the term 
$qQ\xi / {r_+^2}$ and neglecting terms of 
order $\mu \xi^{3/2} (r_+-r_-)^{-1/2}/r_+$ in Eq. (\ref{Eq6}) 
we actually assumed that $\mu b \ll |qQ|$. Thus, we have 
a series of inequalities 
$q/ b^2 = Q/{r_+}^2 \leq 1/Q \ll q/\mu b$, which implies $b \gg \mu$. 
Hence, the lower bound Eq. (\ref{Eq12}) is valid 
for bodies with {\it negligible} self-gravity, which is consistent
with the test particle approximation. In addition, the series of 
inequalities $b \ll r_+ \leq (Q/{r_+}^2)^{-1}=b^2/|q|$ imply $|q|\ll b$.

Assuming the validity of the GSL, one can derive an upper bound to 
the entropy $S$ of an arbitrary system of proper energy $E$ 
and charge $q$:

\begin{equation}\label{Eq13}
S \leq \pi (2E b-q^2)/ \hbar\  .
\end{equation}
It is evident from the minimal
black-hole area increase Eq. (\ref{Eq12}) that in order for the GSL to
be satisfied $[(\Delta S)_{tot} \equiv (\Delta S)_{bh} -S \geq 0]$,
the entropy $S$ of the charged system must be bounded as in
Eq. (\ref{Eq13}). This upper bound is {\it universal} in the sense that it
depends only on the {\it system's} parameters 
(it is {\it independent} of the black-hole parameters $M$
and $Q$).

We emphasized that the universal upper bound Eq. (\ref{Eq13}) is
derived for bodies with {\it negligible} self-gravity. Nevertheless, 
this improved bound is also very appealing from a black-hole physics point
of view: consider a {\it charged} Reissner-Nordstr\"om
black hole of charge $Q$. Let its energy be $E$; then its 
surface area is given by $A=4 \pi {r_{+}}^2 =4\pi (2E r_+-Q^2)$. 
Now since $S_{bh}=A/4\hbar$,
$S_{bh}=\pi (2Er_+-Q^2) /\hbar$, which is the {\it maximal} 
entropy allowed by the upper bound Eq. (\ref{Eq13}). Thus, {\it all}
Reissner-Nordstr\"om black holes {\it saturate} the bound. This proves
that the Schwarzschild black hole is {\it not} unique from a black-hole entropy point of
view, removing the disturbing feature of the entropy bound
Eq. (\ref{Eq2}). This is {\it precisely} the kind of universal upper 
bound we were hoping for !

Evidently, systems with negligible self-gravity (the charged system in our
gedanken experiment) and systems with maximal gravitational effects 
(i.e., charged black holes) both satisfy the upper bound
Eq. (\ref{Eq13}). Therefore, this bound appears to be of universal
validity. Still, it should be recognized that the upper bound
Eq. (\ref{Eq13}) is established only for bodies with negligible
self-gravity. It is of great interest to derive the bound for strongly
gravitating systems. One piece of evidence exist concerning the
validity of the bound for the specific example of a system composed 
of a charged black hole in thermal equilibrium with radiation \cite{Zas2}.

In summary, using a gedanken experiment in which 
an entropy-bearing charged system falls 
into a charged black hole, and assuming the validity of the GSL, one
can derive a {\it universal upper bound} to the entropy of 
a {\it charged} system. An important goal is obviously to clarify the ultimate relation of
the bound to black holes. In fact this relation is reflected in the
numerical factor of $\pi$ which multiply the $q^2$ term. We believe that some
other proof, presumably a more complicated one, could establish this
value of the numerical coefficient. [It should be stressed that 
this is also the current situation for the original
upper bound Eq. (\ref{Eq2}), which was first suggested in the context
of black-hole physics \cite{Beken5}. The relation of the original
bound to black holes is reflected in the numerical factor of $2 \pi$
appearing in it]. Nevertheless, our main goal in this paper was to
prove the {\it general} structure of the universal upper bound for 
{\it charged} systems; the new and interesting observation of this paper is the role
of electric {\it charge} in providing an important {\it limitation} on the 
entropy which a finite physical system can have.

The intriguing feature of our derivation is that it uses a law
whose very meaning stems from gravitation (the GSL, or equivalently
the area-entropy relation for black holes) to derive a universal 
bound which has {\it nothing} to do with 
gravitation [written out fully, the entropy bound 
would involve $\hbar$ and $c$, but {\it not}
$G$]. This provides a striking illustration of the {\it unity} of physics.

\bigskip
\noindent
{\bf ACKNOWLEDGMENTS}
\bigskip

I wish to thank Professor Jacob D. Bekenstein and Avraham E. Mayo for
stimulating discussions. 
This research was supported by a grant from the Israel Science Foundation.


\begin{thebibliography}{99}

\bibitem{Chris} D. Christodoulou, Phys. Rev. Lett. {\bf 25}, 1596 (1970).

\bibitem{ChrisRuf} D. Christodoulou and 
R. Ruffini, Phys. Rev. D {\bf 4}, 3552 (1971).

\bibitem{Haw1} S. W. Hawking, Phys. Rev. Lett. {\bf 26}, 1344 (1971).

\bibitem{Haw2} S. W. Hawking, Commun. Math. Phys. {\bf 25}, 152 (1972).

\bibitem{Car} B. Carter, Nat. Phys. Sci. {\bf 238}, 71 (1972).

\bibitem{BarCarHaw} J. M. Bardeen, B. Carter and S. W. Hawking, 
Commun. Math. Phys. {\bf 31}, 161 (1973).

\bibitem{Beken1} J. D. Bekenstein, Ph.D. thesis, Princeton 
University,1972 (unpublished).

\bibitem{Beken2} J. D. Bekenstein, Lett. Nuov. Cim. {\bf 4}, 737 (1972).

\bibitem{Beken3} J. D. Bekenstein, Phys. Rev. D {\bf 7}, 2333 (1973).

\bibitem{Beken4} J. D. Bekenstein, Phys. Rev. D {\bf 9}, 3292 (1974).

\bibitem{Beken5} J. D. Bekenstein, Phys. Rev. D {\bf 23}, 287 (1981).

\bibitem{Zas1} O. B. Zaslavskii, Phys. Lett. {\bf A 160}, 339 (1991).

\bibitem{Zas2} O. B. Zaslavskii, Gen. Rel. Grav. {\bf 24}, 973 (1992).

\bibitem{Zas3} O. B. Zaslavskii, ``Bekenstein entropy upper bound
 and laws of thermodynamics'', Kharkov University preprint, 1993.

\bibitem{LiLiu} L. X. Li and L. Liu, Phys. Rev. D {\bf 46}, 3296 (1992).

\bibitem{Sorkin} R. D. Sorkin, R. M. Wald and Z. Z. Jiu, Gen. Rel. 
Grav. {\bf 13}, 1127 (1981).

\bibitem{Zurek} W. H. Zurek, and D. N. Page, Phys. Rev. D {\bf 29}, 
628 (1984).

\bibitem{Beken6} J. D. Bekenstein, Phys. Rev. D {\bf 30}, 1669 (1984).

\bibitem{Beken7} J. D. Bekenstein, Phys. Rev. D {\bf 49}, 1912 (1994).

\bibitem{BekenSchi} J. D. Bekenstein and M. Schiffer, Int. J. Mod. Phys. C {\bf 
1}, 355 (1990).

\bibitem{SchiBeken} M. Schiffer and J. D. Bekenstein, Phys. Rev. D {\bf 39}, 110
9 (1989).

\bibitem{BekenGuen} J. D. Bekenstein and E. I. Guendelman, Phys. 
Rev. D {\bf 35}, 716 (1987).

\bibitem{Hod1} S. Hod, Submitted to Phys. Rev. Lett.

\bibitem{Carter} B. Carter, Phys. Rev. {\bf 174}, 1559 (1968).

\bibitem{note1} As our task is to challenge the validity of the GSL in 
the most 'dangerous' situation [with $(\Delta S)_{bh}$ as small as 
possible] and to derive the {\it strongest} bound on entropy, 
we consider the case of a charged body which is 
captured from a radial turning point of its motion. This {\it minimize} the 
increase in black-hole surface area, and thus allows one to derive the 
strongest entropy bound.

\bibitem{Haw3} S. W. Hawking, Nature {\bf 248}, 30 (1974); Commun. Math. Phys. {
\bf 43}, 199 (1975).

\bibitem{note2} As the charged body approaches the horizon it tends to
  be polarized by the black-hole electric field (the same happens to the
  black hole itself). This {\it reduces} the electromagnetic repulsion
  between the two, thus decreasing the threshold energy required in
  order to be captured by the black. Hence, the minimal black-hole area increase
  is even smaller than the one given by Eq. (\ref{Eq11}).

\end{thebibliography}
\end{document}